\documentclass[aps,prc,twocolumn,floatfix]{revtex4}
\usepackage{epsfig}
\usepackage{subfigure}

\begin{document}

\title{Elliptic flow ($v_2$) in $pp$ collisions at energies available
  at the CERN Large Hadron Collider: A hydrodynamical approach}
\author{S. K. Prasad}
\affiliation{Variable Energy Cyclotron 
  Centre, 1/AF Bidhan Nagar, Kolkata 700 064, India}            
\author{Victor Roy}
\affiliation{Variable Energy Cyclotron 
  Centre, 1/AF Bidhan Nagar, Kolkata 700 064, India}            
\author{S. Chattopadhyay}
\affiliation{Variable Energy Cyclotron 
  Centre, 1/AF Bidhan Nagar, Kolkata 700 064, India}            
\author{A. K. Chaudhuri}
\affiliation{Variable Energy Cyclotron 
  Centre, 1/AF Bidhan Nagar, Kolkata 700 064, India}            
\date{\today}

\begin{abstract}
  At Large Hadron Collider energy, the expected large multiplicities suggest the 
  presence of collective behavior even in $pp$ collisions. 
  A hydrodynamical approach has been applied to estimate the expected 
  elliptic flow measured by the azimuthal asymmetry parameter $v_2$ 
  in $pp$ collisions at $\surd$s = 14 TeV. $v_2$ of $\pi^-$ is found to 
  be strongly dependent on the parton density profile inside a proton [e.g., 
  surface diffuseness parameter ($\xi$)]. For $\xi$ = 0.105 fm, $v_2$ is found 
  to be positive and larger compared to that at $\xi$ = 0.25 fm.
  The centrality dependence of $v_2$ has also been studied.
\end{abstract}

\maketitle

\section{Introduction}
The elliptic flow is found to be one of the most interesting observables at 
the relativistic heavy-ion collider (RHIC). 
Large elliptic flow in noncentral Au+Au collisions confirms fluidlike
behavior of the produced matter~\cite{rhic_v2}. Elliptic flow measures 
the momentum anisotropy of the produced particles. 
In noncentral collisions between two identical nuclei, 
the reaction zone is spatially asymmetric. The 
rescattering process among the produced particles 
(locally isotropic in momentum space) transfers 
this spatial asymmetry into the momentum space, 
and the momentum distribution of the produced particles 
becomes anisotropic. Elliptic flow is an early time phenomenon. 
It is a sensitive probe to the: (i) degree of thermalization, 
(ii) equation of state (EOS), and (iii) transport coefficients~\cite{flow1,
  flow2, flow3, Kolb:2001qz, Hirano:2004en}. 

In $pp$ collisions at RHIC energy, the average multiplicity 
$\langle n_{ch} \rangle$ is significantly 
low for any collective effects to be seen, however, detailed 
studies of various observables e.g., Hanbury-Brown-Twiss 
parameters~\cite{mikelisa} and charged particle 
spectra gave hints that even in 
$pp$ collisions, collective models (e.g., hydrodynamics) 
might give a satisfactory explanation for a large part of the
data. Applicability of hydrodynamics in a small system such as $pp$ is
uncertain. Hydrodynamics requires local thermal equilibration, which
can be achieved only if the mean-free path of the constituents is
small compared to the size of the system $\lambda$ $\textless \textless$ R. In
$pp$ collisions, size of the system is not large $\lambda$ $\sim$ R $\sim$ 1
fm. However, in Ref.~\cite{akcplbpp}, it is argued that, if the
medium is isotropized within a time scale $\tau_i$, hydrodynamics may
be applicable beyond $\tau_i$. See Ref.~\cite{akcplbpp} and references
therein, where the applicability of hydrodynamics in central $pp$
collisions after $\tau_i$ $\sim$ 0.2 fm is justified. The scenario of
collective expansion of matter created in $pp$ collisions at Large
Hadron Collider (LHC)
energy is discussed in Ref.~\cite{bozek}.
The expectation that hydrodynamics can be applicable in high-energy
proton-proton collision is not new, for example, 
see Ref.~\cite{larry}. Recently, the possibility to observe
a collective expansion signal - in the form of an azimuthal anisotropy
of particle production with respect to the reaction plane - caused by
multipartonic interactions in proton-proton collisions at the
LHC is studied~\cite{denterria}. In Ref.~\cite{denterria}, different values of integrated
$v_2$ are predicted, which range from -3\% to 10\% depending upon the
profile of matter distribution in transverse space of the colliding
protons, and it is argued that the study of hadron anisotropy with
respect to the reaction plane in $pp$ collisions at LHC energies can
provide important information on the proton shape and
structure. In Ref.~\cite{urs}, large elliptic 
flow $v_2$ $\sim$10-20\% is
predicted in high-multiplicity $pp$ collisions at the LHC if hot-spot-like
structures are produced in the initial collisions.  

In $pp$ collisions at LHC energies, average multiplicity 
will increase, and there could be events with multiplicity 
comparable to the multiplicity in peripheral Au+Au collisions at the RHIC.  
A picture based on two separate transverse distance scales in $pp$ 
collisions at higher energies gives an impact-parameter dependence of 
$pp$ inelastic (INEL) collisions. In the case of central $pp$ collisions, the 
distribution of hard partons ($x \geq 10^{-2}$) in the two 
colliding nucleons will 
overlap, while in large impact-parameter collisions, partons 
with x $<< 10^{-2}$ will overlap with significant 
probability~\cite{PRDFrank}. A trigger on hard dijet 
production can quantitatively distinguish between a 
central and a peripheral collision.  In this scenario, these
high-energy $pp$ collisions can pictorially be represented as AA
collisions, and the impact-parameter dependence of initial 
eccentricity $\varepsilon_x$ can be estimated. Significantly 
large energy density, which generates large multiplicity in $pp$ collisions 
at the LHC will lead to 
rescatterings, thereby creating a system where hydrodynamics can be 
applied and the parameters such as $v_2$ can be estimated. 
Here we report the results obtained by applying the 
hydrodynamic evolution to the system formed in $pp$ 
collisions at the LHC energy ($\surd$s = 14 TeV).\\ 
The paper is organized in the following way. Section II 
describes the details of the code (AZYHYDRO) used for the 
hydrodynamic evolution of the system, and we explain 
how we incorporated the impact-parameter 
dependence of the $pp$ INEL cross section. 
In Sec. III, the results are 
presented and are discussed. We have explained how the initial energy 
density is chosen for the hydrodynamic evolution of the 
system at $\surd$s = 14 TeV. The $p_t$ spectra, mean $p_t$, 
dN/dy, spatial eccentricity, and $v_2$ with their 
dependence on impact parameter b or $p_t$ are presented. In Sec. 
IV, we summarize our results and conclude. 

\section{The Hydrodynamic model}

We briefly describe the hydrodynamic model used to obtain 
space-time evolution of fluid formed in $pp$ collisions. 
Details can be found in Refs.~\cite{hydro1, Kolb:2003dz}.  
The equation of motion of a  relativistic ideal fluid follows from the local 
conservation laws of energy and momentum, and other conserved currents 
(e.g., baryon number),\\
\begin{eqnarray}
  \partial_{\mu}T^{\mu \nu}(x) &=& 0  \label{eq1}\\
  \partial_\mu j^\mu(x)&=&0 \label{eq2}.
\end{eqnarray}
Ideal fluid decompositions of energy-momentum tensor 
($T^{\mu\nu}$) and baryon four-current ($j^\mu$) are as follows:
\begin{eqnarray}
  T^{\mu\nu}(x) &=& [e(x)+p(x)]u^\mu(x)u^\nu(x)-g^{\mu\nu}p(x),\\
  j^{\mu}(x) &=& n(x)u^{\mu}(x),
\end{eqnarray}
\noindent where
$e(x)$ is the energy density, $p(x)$ is the pressure, and $n(x)$ is the conserved
baryon number density at point $x^{\mu} = \{t, x, y, z\}$. $u^{\mu}$ 
is the hydrodynamic four-velocity, $u^{\mu} = \gamma (1, v_x, v_y, v_z)$
with  $\gamma = \frac{1}{\sqrt{(1-v_x^2-v_y^2-v_z^2)}}$. 
The publicly  available AZYHYDRO code solves 2+1-dimensional hydrodynamics. 
At high collision energies, relativistic kinematics and its influence 
on the particle production process implies longitudinal-boost invariance
of the collision fireball near midrapidity~\cite{bjorken}. As a result, 
the longitudinal velocity field scales as $v_z = \frac {z}{t}$, and 
it is convenient to use a coordinate system spanned by longitudinal proper
time $\tau = t\sqrt{1-v_z^2}$ and the space-time rapidity 
$\eta = \frac{1}{2}ln[\frac{t+z}{t-z}]$ instead of t and
z. Longitudinal-boost invariance is then equivalent to $\eta$ independence. 
By assuming the longitudinal-boost invariance, we reduce the number
of energy-momentum conservation equations from four to three, viz.
two transverse and one time. This will restrict us as we see the 
effect of rapidity dependence of transverse flow pattern.\\ 

The initial thermalization stage lies outside the domain of applicability 
of hydrodynamical approach and must be replaced by the initial condition for 
the hydrodynamical evolution.
We assume that the fluid is thermalized at the initial time 
$\tau_i$. We have systematically performed the calculations for three
different values of $\tau_i$ (0.2, 0.4, and 0.6 fm). At the initial 
time, transverse velocity of the fluid is zero $v_x(x,y)=v_y(x,y)=0$. 
At an impact-parameter b, the initial  
energy density is assumed to be distributed as
\begin{equation} 
  \epsilon_i(x,y,{\bf b})= \epsilon_0 N_{coll}(x,y,{\bf b}),
\end{equation}
\noindent where $N_{coll}(x,y,{\bf b}) 
\propto T(x+\frac{b}{2},y)T(x-\frac{b}{2},y)$ 
is a Glauber model calculation for the transverse 
profile of the (partonic) binary collisions at 
impact parameter ${\bf b}$. The central energy 
density $\epsilon_0$ does not depend on the 
impact parameter of the collisions. 

For the Glauber model calculation of collision number 
distribution, we assume a Woods-Saxon profile for the 
(partonic) density distribution of the colliding
protons, 
\begin{equation}
  \rho(r) = \frac{\rho_0}{1+e^{(r-R)/\xi}} 
\end{equation}
with   R=1.05 fm. The calculations are performed at two values of
proton diffuseness parameter (i.e., $\xi$ = 0.105 fm and $\xi$ = 0.25
fm) (see Ref.~\cite{denterria} and references therein) . The value of
$\rho_0$ is fixed in such a way that the total INEL
cross section in $pp$ collisions at a given energy is reproduced in
the Glauber model calculation. The values of $\rho_0$, required to
produce the INEL cross section are 0.45 and 0.70 at $\sqrt{s}$ =
0.2 and 14 TeV, respectively, for $\xi$ = 0.25. The total
INEL cross sections in 
$pp$ collisions at $\sqrt{s}$ = 0.2 and 14 TeV are taken to be
40 and 80 mb, respectively.   
The  parton-proton thickness function is 
given by the optical path length:
\begin{equation}
  T(x,y)=\int^\infty_{-\infty} \rho(x,y,z)dz.
\end{equation}
Hydrodynamical equations Eq.\ref{eq1}, and \ref{eq2} are 
closed only with an EOS $p=p(e,p)$.
We have used the EOS, which is composed of 
lattice EOS and hadron resonance gas EOS. Recently Cheng
$et. al.$~\cite{cheng} presented high statistics lattice 
QCD results for the bulk thermodynamic observables (e.g.,
pressure, energy density, entropy, etc.). We have parametrized
the entropy density as
\begin{equation}
  \frac{s}{T^3} = \alpha + [\beta + \gamma]
  [1 + tanh\frac{T - T_c}{\delta T}].
\end{equation}
The values of the parameters $\alpha, \beta, \gamma$ are chosen in such
a way that the lattice simulation of s/$T^3$ is best fitted~\cite{akcplb}.
We have taken the crossover temperature $T_c$ = 196 MeV. 
From the parametric form of entropy density, pressure and energy
density can be obtained by using the thermodynamic relations:
\begin{eqnarray}
  p(T) = \int s(T^{\prime})dT^{\prime},\\
  \epsilon(T) = Ts - p.
\end{eqnarray}
We complement the lattice EOS~\cite{cheng} by a hadronic resonance 
gas (HRG) EOS, which comprises all the resonances below the mass 2.5 GeV.
The entropy density of complete EOS is obtained as,
\begin{equation}
  s = 0.5[1 + tanh(x)]s_{HRG} + 0.5[1 - tanh(x)]s_{LATTICE}
\end{equation}
with x = $\frac{T - T_c}{\delta T}$, $\delta T$ = 0.1
$T_c$. $s_{HRG}$ is the entropy density for HRG, and
$s_{LATTICE}$ is the entropy density for lattice.

As the system expands, its volume increases, and the density 
decreases, as a result, after some time, the mean-free path of the particles 
becomes larger than the system size, and the 
concept of local thermalization breaks down. Thus, the hydrodynamic 
evolution has to be stopped by applying the freeze-out criteria. 
We assume that freeze out occurs at a fixed temperature $T_F$. 
We have used three different values of $T_F$ 
(130, 140, and 150 MeV) in our calculation. 
By using the standard Cooper-Frye formalism~\cite{CooperFrye}, we
calculate the invariant distribution of particles at 
the freeze-out hypersurface. 
In the Cooper-Frye formalism, 
the invariant  distribution is given by the following equation:
\begin{equation}
  E\frac{dN_i}{d^3p} = 
  \frac{dN_i}{dyp_Tdp_Td\phi} = 
  \frac{g_i}{(2\pi)^3}\int_{\Sigma}{f_i(p.u(x),x)p^{\mu}d^3\sigma_{\mu}},
\end{equation}
where $d^3\sigma_{\mu}$ is the outward normal vector on the 
freeze-out hypersurface $\Sigma$ 
such that $p^{\mu}f_id^3\sigma_{\mu}$ is the local flux 
of particles with the momentum p through this surface, 
and the distribution function is
\begin{equation}
  f_i(E,x)=\frac{1}{exp[\frac{(E-\mu_i(x))}{T(x)}]\pm 1}.
\end{equation}
By applying Lorentz boost, we will get the value of local flow
velocity $u^{\mu}(x)$ to the global reference frame by the substitution
E$\rightarrow p^{\mu}u_{\mu}$, where $\mu_i(x)$ and $T(x)$ are the chemical
potential of particle species $i$ and the local temperature 
along hypersurface $\Sigma$, respectively. 
From the invariant distribution, particle multiplicity ($dN_{ch}/dy$), 
mean $p_T$ ($<p_T>$), elliptic flow ($v_2$), etc.,
can easily be computed.

\section{Results}
\subsection{Estimation of initial energy density in $pp$ collisions at LHC}
\begin{figure}[here]
  \centering
  \includegraphics[width=8.5cm]{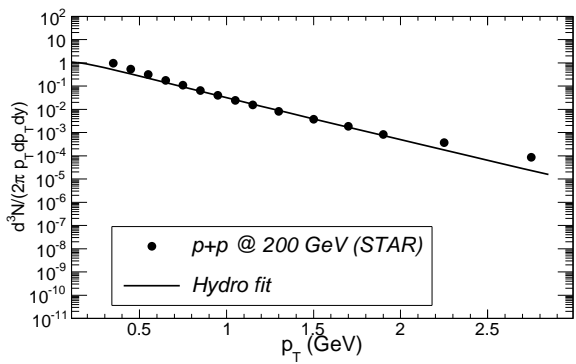}
  \caption{The filled circles are the RHIC data for the transverse momentum 
    distribution for $\pi^-$ in $pp$ collisions at $\sqrt s$ = 200 GeV.
    The solid line is the hydrodynamical model fit to the data at
    diffuseness parameter $\xi$ = 0.25 fm.}
  \label{fig1}
\end{figure}

\begin{figure}[here]
  \centering
  \includegraphics[width=8.5cm]{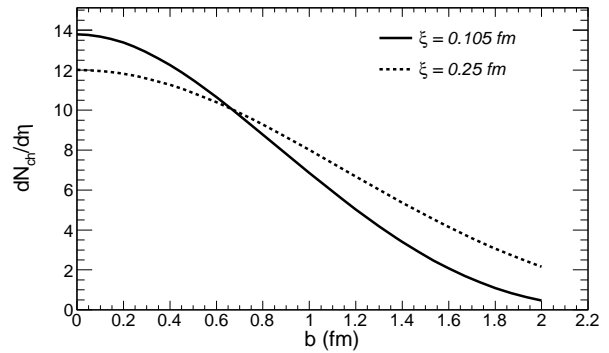}
  \caption{Variation of d$N_{ch}$/dy with impact parameter b, 
    for $\sqrt{s}$ = 14 TeV for $\xi$ = 0.25 fm (dotted line) and
    $\xi$ = 0.105 fm (solid line).}
  \label{dndyb}
\end{figure}

\begin{figure}[ht]
  \centering
  \includegraphics[width=8.5cm ]{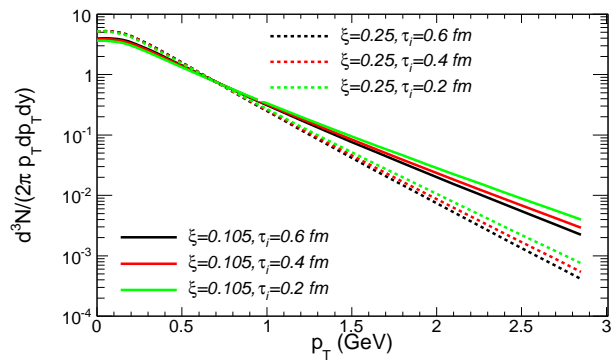}
  \caption{(Color online)The minimum bias $p_T$ spectra for three initial times 
    ($\tau_i$ = 0.2, 0.4, 0.6 fm),
    with $\xi$ = 0.25 fm (dotted lines) and $\xi$ = 0.105 fm (solid
    lines) at $\sqrt{s}$ = 14 TeV.}
  \label{ptspectraTi}
\end{figure}

\begin{figure}[ht]
  \centering
  \includegraphics[width=8.5cm]{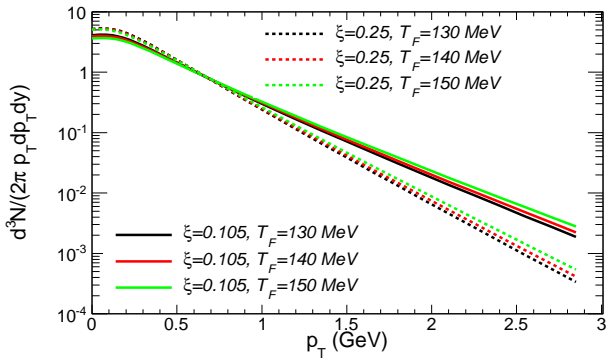}
  \caption{(Color online) The minimum bias $p_T$ spectra for three freeze-out temperatures 
    ($T_F$ = 130, 140, 150 MeV)
    with $\xi$ = 0.25 fm (dotted lines) and $\xi$ = 0.105 fm (solid
    lines) at $\sqrt{s}$ = 14 TeV.}
  \label{ptspectraTf}
\end{figure}

In the absence of experimental guidance in $pp$ collisions at 
the LHC ($\sqrt{s}$ = 14 TeV), 
it is difficult to fix the initial energy density $\epsilon_i$ 
for hydrodynamical calculations in $pp$ collisions at the LHC. 
We choose the initial energy density such that the experimentally
measured or extrapolated values of $dN_{ch}/d\eta$ in $pp$ collisions
at given $\sqrt{s}$ is reproduced. We have fitted the experimentally
measured $p_{T}$ spectra of $\pi^-$ in $pp$ collisions at $\sqrt{s}$
= 200 GeV from the STAR experiment at the RHIC, and they are shown in Fig.\ref{fig1}.
The filled circles are the RHIC data points for the transverse momentum 
distribution of $\pi^-$. The solid line is the hydrodynamical model
fit to the data with proton 
diffuseness parameter $\xi$ = 0.25 fm.
Note that the $pp$ collisions are measured at minimum bias, 
without any centrality or 
impact-parameter dependence. 
To compare the experimental 
data with hydrodynamic simulations, we compute the minimum bias spectra as
\begin{equation}
  \frac{dN}{dyd^2p_T}=\frac{\int_0^{b_{max}} 2\pi b 
    \frac{dN(b)}{dyd^2p_T} db} {\int_0^{b_{max}} 2\pi b   db}
\end{equation}
\noindent where $\frac{dN(b)}{dyd^2p_T}$ is the $\pi^-$ 
invariant distribution at an impact parameter b. 
We have integrated upto $b_{max}$ = 1.6 fm, which covers 
$\sim$ 90\% of thr INEL cross section.
Generally, one does not apply 
hydrodynamics in $pp$ collisions. The system size possibly is too small for 
macroscopic concepts to be valid. However, in Ref.~\cite{larry}, there are
some discussions on applications of hydrodynamical calculations to
such systems. We observe that the $p_{T}$ spectra as measured 
by the STAR Collaboration at the RHIC ($\sqrt{s}$ = 200 GeV) 
for $pp$ collisions is reasonably well
explained in the hydrodynamic model. 
In the low $p_T$ region $p_T <$ 0.5 GeV/c,  the simulation underestimates the 
data because a large fraction of low  $p_T$ pions are probably coming from 
various resonance decays, which are not considered in the simulation. 
Agreement with data in the low $p_T$ region will improve if 
resonance contributions are accounted for. Hydrodynamical 
simulation also underestimates the data at high $p_T> $ 2 GeV/c, 
which indicates departure from ideal hydrodynamics because of dissipative effects. 
Also, at large $p_T$, sources other than hydrodynamics contribute to 
particle production.

Charged-particle pseudorapidity
distribution in the central rapidity region in p+p and p+$\bar{p}$ 
interactions is expected to have a power-law dependence 
on center-of-mass energy~\cite{alicepaper1}. Therefore, 
we use the relation,
\begin{equation}
  \left. \frac{dN_{ch}}{dy}\right|_{LHC} = a\times
  (\sqrt{s}_{LHC})^b,
  \label{dec}
\end{equation}
\noindent where $a$ and $b$ are the parameters whose values are
obtained as $a$ = 0.7166 and $b$ = 0.2171 by fitting the
$\sqrt{s}$ dependence of INEL charged-particle  
pseudorapidity density in the
central rapidity region in p+p and p+$\bar{p}$ 
interactions~\cite{alicepaper1}. The extrapolated minimum bias
charged particles pseudorapidity density at y = 0 at the LHC ($\sqrt{s}$
= 14 TeV) is determined to be 5.69.
For performing simulations at the LHC, the initial energy 
density is fixed accordingly.
The energy density required to reproduce the extrapolated
$dN_{ch}/d\eta$ 
is different for different values of initial time ($\tau_i$) and 
freeze-out temperature ($T_{F}$).

\subsection{Mean multiplicity (dN/dy), mean $p_t$ ($<p_t>$) and $p_t$ spectra}
The parameters that are used in this calculation for $pp$ collisions need
to be varied as these parameters can be fixed only after detailed
experimental investigations. Therefore, we perform our
calculations for three different values of freeze-out  
temperature ($T_F$ = 130, 140, and 150 MeV) and three different
values of initial 
time ($\tau_i$ = 0.2, 0.4, and 0.6 fm) with two values of
diffuseness parameter 
($\xi$ = 0.25 fm and $\xi$ = 0.105 fm). 

The initial energy density ($\epsilon_i$), the minimum 
bias average multiplicity ($\frac{dN_{ch}}{dy}$) 
at y = 0, the mean $p_T$, and the $p_T$ integrated $v_2$ in $pp$
collisions at LHC energy ($\sqrt{s}$ = 14 TeV), are calculated
and are summarized in Tables~\ref{table1} and~\ref{table2} for
various initial conditions at two values of diffuseness parameters. 

\begin{table}[h]\footnotesize
  \caption{The minimum bias multiplicity $dN_{ch}$/dy, the mean $p_T$
    and the $p_T$ integrated $v_2$ for various freeze-out temperatures
    for both diffuseness parameters at a fixed $\tau_i$.}
  \label{table1}
  \begin{tabular}{|c|c|c|c|c|c|c|}
    \hline
    $\tau_i$(fm)        & $\xi$(fm)     & $T_F$(MeV)           &
    $\epsilon_i$ (GeV/$fm^3$)  & $dN_{ch}/dy$ & \textless $p_T$
    \textgreater(GeV/c)  & $v_2$(\%) \\ 
    \hline 
    0.6  & 0.25 & 130    & 20.7  &  5.68  & 0.56 & 0.34 \\
    0.6  & 0.25 & 140    & 26.2  &  5.68  & 0.57 & 0.37 \\
    0.6  & 0.25 & 150    & 35.0  &  5.68  & 0.59 & 0.40 \\
    \hline
    0.6  & 0.105 & 130  & 31.4  & 5.69   &  0.68 & 1.38 \\
    0.6  & 0.105 & 140  & 40.0  & 5.69   &  0.70 & 1.38 \\
    0.6  & 0.105 & 150  & 53.8  & 5.69   &  0.73 & 1.40 \\
    \hline
  \end{tabular}
\end{table}

\begin{table}[h]\footnotesize
  \caption{The minimum bias multiplicity $dN_{ch}$/dy, the mean $p_T$
    and the $p_T$ integrated $v_2$ for various initial times
    for both diffuseness parameters at a fixed $T_F$}
  \label{table2}
  \begin{tabular}{|c|c|c|c|c|c|c|}
    \hline
    $T_F$(MeV)    & $\xi$ (fm)  & $\tau_i$(fm)         &
    $\epsilon_i$ (GeV/$fm^3$)  & $dN_{ch}/dy$ & \textless $p_T$
    \textgreater(GeV/c)  & $v_2$(\%) \\ 
    \hline 
    140  & 0.25 & 0.2    & 99.7  &  5.69  & 0.60 & 0.34 \\
    140  & 0.25 & 0.4    & 42.5  &  5.68  & 0.59 & 0.35 \\
    140  & 0.25 & 0.6    & 26.2  &  5.68  & 0.57 & 0.37 \\
    \hline
    140  & 0.105 & 0.2  & 160.7 & 5.69  & 0.76 & 1.32 \\
    140  & 0.105 & 0.4  & 66.5   & 5.69  & 0.72 & 1.30 \\
    140  & 0.105 & 0.6  & 40.0   & 5.69  & 0.70 & 1.38 \\
    \hline
  \end{tabular}
\end{table}

At a fixed initial time ($\tau_i$ = 0.6 fm), when the freeze-out
temperature is increased from 130 to 150 MeV, the mean $p_T$ is
found to increase by 5.3\%  and 7.3\% whereas the $p_T$ integrated $v_2$ is
found to increase by 17.6\% and 1.4\% for $\xi$ = 0.25 and 0.105 fm,
respectively. At a fixed freeze-out temperature ($T_F$ = 140 MeV), when the initial
time is increased from 0.2 to 0.6 fm, the mean $p_T$ is found to
decrease by 5.0\%  and 7.8\%,  whereas the $p_T$ integrated $v_2$ is
found to increase by 8.8\% and 4.5\% for $\xi$ = 0.25 and 0.105 fm,
respectively.  
However, the mean $p_T$ is however, found to be consistently higher for 
the case with 
$\xi$ = 0.105 fm as compared to the case with $\xi$ = 0.25 fm. The
reason is understood. For the sharp proton density distribution
($\xi$ = 0.105 fm), fluid is initialized with higher energy density as
compared to that when the surface is more diffused ($\xi$ = 0.25
fm). Increased transverse pressure in the fluid then leads to
increased mean $p_T$.  

\begin{figure}[here]
 \centering
 \includegraphics[width=8.5cm]{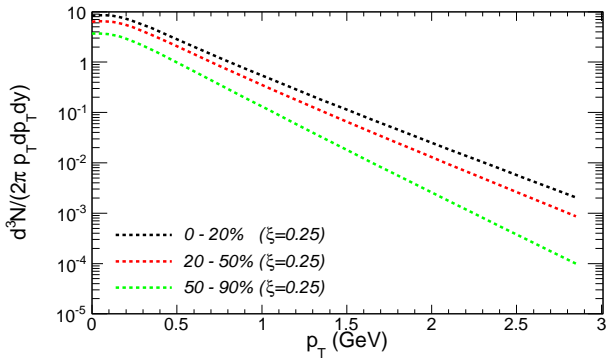}
 \caption{(Color online)The $p_T$ spectra for three different
   centralities with  $\xi$ = 0.25 fm.}
 \label{ptbp25}
\end{figure}

\begin{figure}[here]
 \centering
 \includegraphics[width=8.5cm]{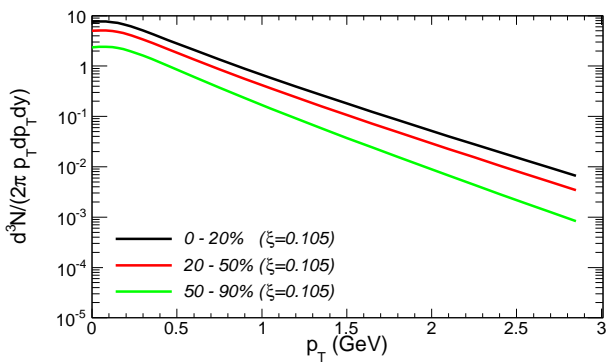}
 \caption{(Color online)The $p_T$ spectra for three different
   centralities with $\xi$ = 0.105 fm.}
 \label{ptbp105}
\end{figure}

In Fig.\ref{dndyb}, we have shown the impact-parameter dependence of 
$dN_{ch}$/dy at y = 0 for both values of diffuseness parameter
$\xi$ = 0.105 fm and $\xi$ = 0.25 fm. As expected, charged-particle
rapidity density decreases with increasing impact parameters for both
$\xi$. This indicates that higher multiplicity 
events in $pp$ collisions are obtained at low-impact parameters,
and a centrality trigger based on multiplicity might select these
interesting events. The rapidity density of charged particles at
mid rapidity for b = 0 is 13.79 for $\xi$ = 0.105 fm and 12.01 for $\xi$
= 0.25 fm. The impact-parameter dependence of rapidity density of a
charged particle for two diffuseness parameters shows an interesting
feature, that for $\xi$ = 0.105 fm, the rapidity density of charged
particles reduces faster as their configuration is close to the hard
sphere; and above some impact parameter, the overlap region reduces
faster. The result is according to our expectation. In peripheral
collisions, the overlap region is comparatively large when the surface
is more diffused. The impact parameter dependance of rapidity
density of charged particles can be predicted to provide size
information on the diffuseness parameter of the colliding protons.

The minimum bias $p_T$ spectra of charged particles for different
values of initial times and freeze-out temperature are shown in
Figs.\ref{ptspectraTi} and\ref{ptspectraTf}, respectively.
Minimum bias $p_T$ spectra are obtained 
by integrating up to $b_{max}$ = 1.6 fm, 
which, as previously indicated, covers $\sim$90\% of INEL cross section. 
It is observed that the $p_T$ spectra are not affected much because of the
change in freeze-out temperature and in initial time for a
particular diffuseness parameter. However, the spectra are different
for different diffuseness parameters. Note that, whenever there is a
change in either $T_F$ at fixed $\tau_i$ or in $\tau_i$ at fixed
$T_F$, the initial energy density ($\epsilon_i$) is adjusted
accordingly so that the average multiplicity ($dN_{ch}/dy$) is always fixed. 

\begin{figure}[here]
  \centering 
  \includegraphics[width=8.5cm]{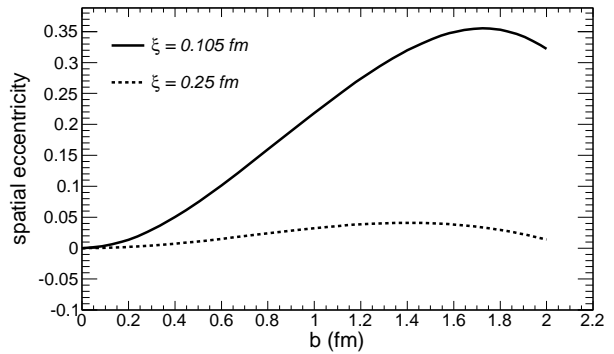}
  \caption{Initial spatial eccentricity, $\varepsilon_x$ as a function of 
    impact parameter for $pp$ collisions at $\sqrt{s}$ = 14 TeV
    for $\xi$ = 0.25 fm (dotted line) and  $\xi$ = 0.105 fm 
    (solid line).}
  \label{fig7}
\end{figure} 

As discussed earlier, the pseudorapidity density of
the cgarged particle is 13.79 and 12.01 for $\xi$ = 0.105 fm and $\xi$ =
0.25 fm, respectively, at b = 0. This suggests that one can possibly
determine two to three centralities experimentally on the basis of
multiplicity. Therefore, we have studied the centrality (or impact
parameter) dependence of 
$p_T$ spectra, $v_2$ etc. for three centralities viz. 0 to 20\%, 20 to
50\% and 50 to 90\%.
The centrality dependence of $p_T$ spectra for
three centralities is
shown in  Figs.\ref{ptbp25} and\ref{ptbp105} for $\xi$ = 0.25 fm and
$\xi$ = 0.105 fm, respectively. As indicated previously, we have neglected
the resonance contribution. Resonance mainly contributes at the low $p_T$
region. Centrality dependence of $p_T$ spectra is qualitatively similar
to that obtained in Au+Au collisions at the RHIC at both $\xi$. As the
collisions become 
more and more peripheral, the slope of the spectra gets steeper,
which indicates reduced source temperature. The result is consistent with
our expectation. Although the spectra looks similar for both $\xi$,
they differ in details (e.g., slope, total yield, etc.). 

\subsection{Elliptic flow ($v_2$)} 
The particle azimuthal distribution can be constructed from different 
quantities such as transverse momentum, multiplicity, or transverse energy
in relatively narrow (pseudo)-rapidity windows~\cite{voloshin}. For
the non zero
impact parameter, the azimuthal distribution of the particles in the reaction
plane is not symmetric (anisotropic) in $\phi$. 
We can decompose the distribution in terms of Fourier
expansion as
\begin{equation}
  E\frac{d^3N}{d^3p} = \frac{d^2N}{2\pi p_tdp_tdy}\times \{ \sum (1 + 2v_ncos(n\phi^{\prime})\},
\end{equation}
\noindent 
where $\phi^{\prime}$ is the angle of azimuth of the outgoing particle with
respect to the reaction plane. The coefficient of the second 
term in the Fourier expansion is
known as the elliptic flow $v_2$. The nonzero $v_2$ describes the 
eccentricity of an ellipselike distribution. The origin of the 
measured anisotropy could be different: hydrodynamical flow, 
shadowing effect, both, etc. What they
have in common is some collective behavior in the 
evolution of the multiparticle 
production process~\cite {voloshin}.

The initial spatial eccentricity of the reaction zone in the 
transverse plane  changes with a change in the impact parameter. 
The $v_2$ is sensitive to this change in spatial eccentricity
and, consequently, to the change in the impact parameter. 
Initial spatial eccentricity $\varepsilon_x$ is defined as
\begin{equation}
  \varepsilon_x(b)  = \frac{\langle y^2 - x^2 \rangle}
  {\langle y^2 + x^2 \rangle},
\end{equation} 
where the angular brackets denote energy density weighted  
averages at the initial time $\tau_i$. In Fig.\ref{fig7}, 
initial eccentricity
$\varepsilon_x$, for the two values of diffuseness parameter $\xi$ = 0.25 fm  
and $\xi$ = 0.105 fm, are shown  as a function of the impact parameter. 
For diffused protons  ($\xi$ = 0.25 fm), 
the spatial eccentricity does not grow significantly.  
However, for the small diffuseness parameter, $\xi$ = 0.105 fm, 
the spatial eccentricity 
$\varepsilon_x$ increases with the impact parameter and reaches a maximum at 
midcentral collisions. The behavior is similar to that of Au+Au collisions.

\begin{figure}[here]
  \centering
  \includegraphics[width=8.5cm]{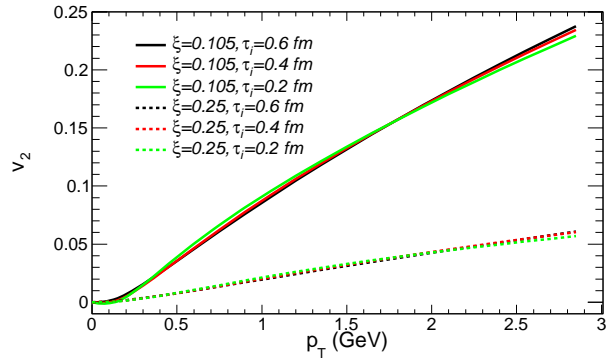}
  \caption{(Color online) The minimum bias differential elliptic flow $v_2$($p_{T}$)
    for three initial times ($\tau_i$ = 0.2, 0.4, 0.6 fm)
    for $\xi$ = 0.25 fm (dotted lines) and $\xi$ = 0.105 fm (solid lines).}.
  \label{v2ptTi}
\end{figure}

\begin{figure}[here]
  \centering
  \includegraphics[width=8.5cm]{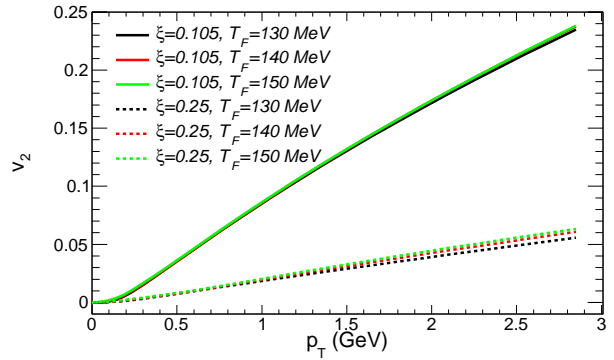}
  \caption{(Color online) The minimum bias differential elliptic flow $v_2$($p_{T}$)
    for three freeze-out temperatures ($T_F$ = 130, 140, 150 MeV)
    for $\xi$ = 0.25 fm (dotted lines) and $\xi$ = 0.105 fm (solid lines).}
  \label{v2ptTf}
\end{figure}

The effect of varying initial conditions ($\tau_i$ and $T_F$) on
minimum bias differential elliptic flow ($v_2$($p_T$))
is studied, and the results are depicted in Figs.~\ref{v2ptTi}
and\ref{v2ptTf}. In Fig.~\ref{v2ptTi}, we have plotted the
minimum bias differential elliptic
flow $v_2 (p_{T})$ for three values of initial time ($\tau_i$ =
0.2, 0.4, 0.6 fm) for both $\xi$.
In Fig.~\ref{v2ptTf}, we have plotted the minimum bias differential elliptic
flow $v_2 (p_{T})$ for three values of freeze-out temperature ($T_{F}$ =
130, 140, 150 MeV) for both $\xi$.

The minimum bias differential elliptic flow [$v_2(p_T)$], is found
to be positive for both $\xi$. The change in
$v_2 (p_{T})$ caused by the change in initial conditions ($\tau_i$ or $T_F$)
is negligible. However, $v_2(p_T)$ 
is found to be very large (reaches up to 24\% at $p_T$ = 3 GeV/c) for
the case with $\xi$ = 0.105 fm, whereas
for $\xi$ = 0.25 fm, it reaches a maximum value up to
6\% (at $p_T$ = 3 GeV/c).

\begin{figure}[here]
  \centering
  \includegraphics[width=8.5cm]{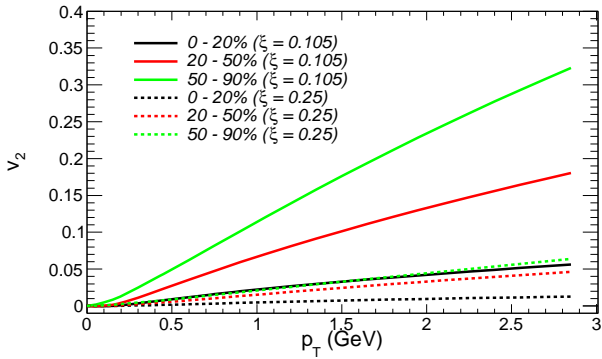}
  \caption{(Color online) The differential elliptic flow
    $v_2(p_T)$ for three different centralities for $\xi$ = 0.25 fm
    (dotted line) and $\xi$ = 0.105 fm (solid line).}
  \label{v2b}
\end{figure}

We have also studied the centrality dependence of $p_T$ integrated
$v_2$ and differential 
elliptic flow $v_2(p_T)$ for both $\xi$, 
for $\tau_i$ = 0.6 fm and $T_F$ = 140 MeV.
The $v_2(p_T)$ for three different
centralities is shown in Fig.~\ref{v2b}. The differential elliptic
flow $v_2(p_T)$ is found to increase when we go toward mid
peripheral, which reaches a maximum upto 33\% (at $p_T$ = 3 GeV/c) for
$\xi$ = 0.105 fm whereas it reaches a maximum value of 7\% for $\xi$ =
0.25 fm. The centrality dependence of $v_2 (p_T)$ is similar for
both $\xi$ apart from the differences in their respective values.

The dependence of mean $p_{T}$, integrated $v_2$, and d$N_{ch}$/dy
with changing the centrality is summarized in Tables~\ref{table3}
and~\ref{table4} for $\xi$ = 0.25 fm and $\xi$ = 0.105 fm,
respectively.

\begin{table}[h]\footnotesize
  \caption{The minimum bias multiplicity $dN_{ch}$/dy, the mean $p_T$,
    and the $p_T$ integrated $v_2$ for three different centralities
    at fixed $\tau_i$ (0.6 fm), $T_{F}$ (140 MeV) and $\xi$ (0.25 fm).}
  \label{table3}
  \begin{tabular}{|c|c|c|c|}
    \hline
    centrality & $dN_{ch}/dy$ & \textless $p_T$ \textgreater(GeV/c) &
    $v_2$(\%)\\
    \hline
    0-20\%    & 10.78  & 0.64 & 0.13 \\
    20-50\%  & 7.55    & 0.60 & 0.34 \\
    50-90\%  & 3.45    & 0.53 & 0.29 \\
    \hline
  \end{tabular}
\end{table}

\begin{table}[h]\footnotesize
  \caption{The minimum bias multiplicity $dN_{ch}$/dy, the mean $p_T$,
    and the $p_T$ integrated $v_2$ for three different centralities
    at fixed $\tau_i$ (0.6 fm), $T_{F}$ (140 MeV) and $\xi$ (.105 fm).}
  \label{table4}
  \begin{tabular}{|c|c|c|c|}
    \hline
    centrality & $dN_{ch}/dy$ & \textless $p_T$ \textgreater(GeV/c) &
    $v_2$(\%)\\
    \hline
    0-20\%    & 11.74  & 0.74 & 0.66 \\
    20-50\%  & 7.47    & 0.71 & 1.34 \\
    50-90\%  & 3.26    & 0.66 & 1.23 \\
    \hline
  \end{tabular}
\end{table}

We observe that the $p_T$ integrated $v_2$ 
shows centrality dependence as 
expected in a hydrodynamic model (i.e., increases 
when we go from central to midperipheral), however, the value of $p_T$
integrated $v_2$
is very small. For midcentral collisions, it reaches 
up to 1.34\% and 0.34\% for $\xi$ = 0.105 fm and $\xi$ = 0.25 fm,
respectively. 

\section{Summary and conclusion}

 To summarize, we have presented a hydrodynamical model study for $pp$ 
 collisions at LHC ($\sqrt{s}$ = 14 TeV) energy.  
 The initial energy density for the hydrodynamic calculation
 is obtained by requiring a condition that the extrapolated rapidity
 density of charged particles 
 should be reproduced in the model. After studying the 
 basic properties of particle production (e.g., spectra, dN/dy, 
 $\langle p_T \rangle$, and their centrality dependence), we have 
 extracted $v_2$ and its dependence on $p_t$ and centralities. As the 
 initial spatial eccentricity depends strongly on the shape of partonic 
 density inside the proton, we have studied two cases of surface diffuseness 
 parameters $\xi$ in Woods-Saxon profile. 
 We have studied the effect of changing initial time $\tau_i$ and 
 freeze-out temperature $T_F$ on elliptic flow. It is found that
 the effect is negligible. The minimum bias differential elliptic flow $v_2(p_T)$
 is found to be very large and reaches up to 24\% for the case with diffuseness 
 parameter, $\xi$ = 0.105 fm. When the diffuseness parameter is 
 0.25 fm, the minimum bias differential elliptic flow is found to
 reach up to
 6\%.
 At LHC energy, the energy density and particle multiplicity are 
 expected to be substantially large, which suggests the creation of a system 
 that shows collective properties. First results from the
 LHC~\cite{alicepaper2, alicepaper3} have already shown charged 
 multiplicity at the midrapidity region going up to 60. Therefore,
 we suggest to measure $v_2$  
 and its centrality dependence in $pp$ collisions at the LHC. The 
 centrality dependence can be studied by selecting events with 
 different multiplicity bins. The estimation of 
 $v_2$ at LHC energy might be affected by the presence of nonflow effects 
 (e.g. jets), but methods based on a scalar product, 
 a cumulant or a standard event 
 plane with a large $\eta$-gap can be applied. The centrality 
 dependence of $v_2$ can be used to obtain the surface diffuseness 
 parameter of protons in $pp$ collisions.

\end{document}